\documentclass{iopart}
\usepackage[pctex32]{graphicx}

\begin{document}

\title{The application of the modified form of B{\aa}th's law to the North Anatolian Fault Zone}
\author{S E Yalcin, M L Kurnaz}
\address{ Department of Physics, Bogazici
University , $34342$ Bebek, Istanbul, Turkey}
\ead{sibelebruyalcin@yahoo.com}
\ead{kurnaz@boun.edu.tr}

\begin{abstract}
Earthquakes and aftershock sequences follow several empirical
scaling laws: One of these laws is B{\aa}th's law for the magnitude
of the largest aftershock. In this work, Modified Form of
B{\aa}th's Law and its application to KOERI data have been studied.
B{\aa}th's law states that the differences in magnitudes between
mainshocks and their largest detected aftershocks are approximately
constant, independent of the magnitudes of mainshocks and it is
about $1.2$. In the modified form of B{\aa}th's law for a given
mainshock we get the inferred largest aftershock of this mainshock
by using an extrapolation of the Gutenberg-Richter
frequency-magnitude statistics of the aftershock sequence. To test
the applicability of this modified law, $6$ large earthquakes that
occurred in Turkey between $1950$ and $2004$ with magnitudes equal
to or greater than $6.9$ have been considered. These
earthquakes take place on the North Anatolian Fault Zone.
Additionally, in this study the partitioning of energy during a
mainshock-aftershock sequence was also calculated in two different
ways. It is shown that most of the energy is released in the
mainshock. The constancy of the differences in magnitudes between
mainshocks and their largest aftershocks is an indication of
scale-invariant behavior of aftershock sequences.
\end{abstract}


\section{INTRODUCTION\protect\\ }
\label{sec:level1}

An earthquake is a sudden and sometimes catastrophic movement of a
part of the Earth's surface \cite{kanamori}. It is caused by the
release of stress accumulated along geologic faults or by volcanic
activity, hence the earthquakes are the Earth's natural means of
releasing stress. When the Earth's plates move against each other,
stress is put on the lithosphere. When this stress is strong enough,
the lithosphere breaks or shifts. As the plates move they put forces
on themselves and each other. When the force is large enough, the
crust is forced to break. When the break occurs, the stress is
released as energy which moves through the Earth in the form of
waves, which we feel and call an earthquake.

Aftershocks are earthquakes in the same region of the
mainshock\cite{kanamori}. Smaller earthquakes often occur in the
immediate area of the main earthquake until the entire surface has
reached equilibrium of stress. There are several scaling laws that
describe the statistical properties of aftershock sequences \cite
{kisslinger,turcotte,shcherbakov}. Gutenberg-Richter
frequency-magnitude scaling law is widely known by seismologists and
scientists. On the Richter scale, the magnitude (M) of an earthquake
is proportional to the log of the maximum amplitude of the earth's
motion. What this means is that if the Earth moves one millimeter in
a magnitude $2.0$ earthquake, it will move ten millimeters in a
magnitude $3.0$ earthquake, $100$ millimeters in a magnitude $4.0$
earthquake and ten meters in a magnitude $6.0$ earthquake. So, the
amplitude of the waves increases by powers of ten in relation to the
Richter magnitude numbers. Therefore, if we hear about a magnitude
$8.0$ earthquake and a magnitude $4$ earthquake, we know that the
ground is moving $10,000$ times more in the magnitude $8.0$
earthquake than in the magnitude $4.0$ earthquake. Numbers for the
Richter scale range from $0$ to $9.0$, though no real upper limit
exists. The difference in energies is even greater. For each factor
of ten in amplitude, the energy grows by a factor of $32$. When
seismologists started measuring the magnitudes of earthquakes, they
found that there were a lot more small earthquakes than large ones.
Seismologists have found that the number of earthquakes is
proportional to $10^{-bM}$. They call this law ``The
Gutenberg-Richter Law" \cite{turcotte, shcherbakov, gutenberg}.

In seismological studies, the Omori law, proposed by Omori in
$1894$, is one of the few basic empirical laws \cite{omori}. This
law describes the decay of aftershock activity with time. Omori Law
and its modified forms have been used widely as a fundamental tool
for studying aftershocks \cite {sornette}. Omori published his work
on the aftershocks of earthquakes, in which he stated that
aftershock frequency decreases by roughly the reciprocal of time
after the main shock. An extension of the modified Omori's law is
the epidemic type of aftershock sequences (ETAS) model \cite
{sornette}. It is a stochastic version of the modified Omori law. In
the ETAS model, the rate of aftershock occurrence is an effect of
combined rates of all secondary aftershock subsequences produced by
each aftershock \cite{knopoff, ogata}.

The third scaling law relating the aftershocks is B{\aa}th's law.
The empirical B{\aa}th's law states that the difference in magnitude
between a mainshock and its largest aftershock is constant,
regardless of the mainshock magnitude and it is about $1.2$
\cite{turcotte, shcherbakov, sornette, bath}. That is
\begin{equation}\label{denklem3}
\Delta m=m_{ms}-m_{as}^{max}
\end{equation}
with $m_{ms}$ the magnitude of the mainshock, $m_{as}^{max}$  the
magnitude of the largest detected aftershock, and $\Delta m$
approximately a constant.

In this article we study the modified form of B{\aa}th's law
\cite{turcotte,shcherbakov}. To study the aftershock sequence in
the North Anatolian Fault Zone (NAFZ) we get the largest aftershock
from an extrapolation of the G-R frequency-magnitude scaling of all
measured aftershocks. We test the applicability of B{\aa}th's law
for $6$ large earthquakes on the North Anatolian Fault Zone (NAFZ).
The emprical form of B{\aa}th's law states that the difference
magnitude between a mainshock and its largest aftershock is
constant, independent of the magnitudes of mainshocks. We also
analyze the partitioning of energy during a mainshock-aftershock
sequence and its relation to the modified B{\aa}th's law.

\section{B{\AA}TH'S LAW AND ITS MODIFIED FORM\protect\\}
\label{sec:level2} B{\aa}th's law states that the differences in
magnitudes between mainshocks and their largest aftershocks are
approximately constant, independent of the magnitudes of mainshocks.
In modified form of B\aa th's law for a given mainshock we get the
inferred largest aftershock of this mainshock by using an
extrapolation of the Gutenberg-Richter frequency-magnitude
statistics of the aftershock sequence. The size distribution of
earthquakes has been found to show a power law behavior. Gutenberg
and Richter, introduced the common description of the frequency of
earthquakes: \cite{gutenberg}
\begin{equation}\label{denklem4}
log_{10}N(\geq m)=a-bm
\end{equation}
where $N(\geq m)$ is the cumulative number of earthquakes with
magnitudes greater than m occurring in a specified area and time
window. On this equation a and b are constants. This relation is
valid for earthquakes with magnitudes above some lower cutoff
$m_{c}$. Earlier studies \cite{turcotte, shcherbakov, frolich} gave
an estimate for this ``b" value between $0.8$ and $1.2$. 
The constant ``a" shows the regional level of
seismicity and gives the logarithm of the number of earthquakes with
magnitudes greater than zero \cite{turcotte, shcherbakov}. In our
analysis a-value is in the range $3.8<a<6.5$. Aftershocks related
with a mainshock also satisfy G-R scaling (\ref{denklem4}) to a good
approximation \cite{turcotte}. In this case $N(\geq m)$ is the
cumulative number of aftershocks of a given mainshock with
magnitudes greater than m. We offer to extrapolate G-R scaling
(\ref{denklem4}) for aftershocks. Our aim is to obtain an upper
cutoff magnitude in a given aftershock sequence. We find the
magnitude of this inferred ``largest" aftershock $m^{*}$ by formally
taking $N(\geq m^{*})=1$ for a given aftershock sequence.
\begin{equation}\label{denklem5}
a=bm^{*}
\end{equation}
This extrapolated  $m^{*}$ value will have a mean value and a
standard deviation from the mean value. We apply the B{\aa}th's law
to the inferred values of $m^{*}$ and then, we can write
\begin{equation}\label{denklem6}
\Delta m^{*}=m_{ms}-m^{*}
\end{equation}
where $m_{ms}$ is the magnitude of the mainshock and $\Delta m^{*}$
is approximately a constant. Substitution of equations
(\ref{denklem5}) and (\ref{denklem6}) into equation (\ref{denklem4})
gives
\begin{equation}\label{denklem7}
log_{10}[N(\geq m)]=b(m_{ms}-\Delta m^{*}-m)
\end{equation}
 with b, $m_{ms}$, and $\Delta m^{*}$ specified, the frequency-magnitude distribution of
aftershocks can be determined using equation (\ref{denklem7}). In
extrapolating the G-R scaling (\ref{denklem4}) the slope of this
scaling or b-value plays an important role in estimating the largest
inferred magnitude $m^{*}$.

\section{APPLICATION OF THE MODIFIED FORM OF B{\AA}TH'S LAW TO THE
NORTH ANATOLIAN FAULT ZONE (NAFZ)} \label{sec:level3} We applied
modified form of B{\aa}th's law by considering $6$ large earthquakes
on the NAFZ. These earthquakes occurred between $1950$ and $2004$. The
data were provided by Bogazici University Kandilli Observatory
and Earthquake Research Institute (KOERI)\cite{kandilli}. The $6$
earthquakes considered had magnitudes $m_{ms}\geq 6.9$. The
important point is that they were sufficiently separated in space
and time so that no aftershock sequences overlapped with other
mainshocks. Earthquakes form a hierarchical structure in space and
time. Therefore, in some cases it is possible to discriminate
foreshocks, mainshocks, and aftershocks. But, generally this
classification is not well defined and can be ambiguous. One of our
main problems in the study of aftershocks is to identify what is and
what is not an aftershock \cite{molchan}. To specify aftershocks we
defined space and time windows for each sequence. In each case we
consider a square area centered on the mainshock epicenter. The
linear size of the box is taken to be of the order of the linear
extent of the aftershock zone L, which scales with the magnitude of
the mainshock as
\begin{equation}\label{denklem8}
L=0.02\times 10^{0.5 m_{ms}} km
\end{equation}
This equation was given by Yan Y. Kagan \cite{kagan}. Previously,
Shcherbakov and Turcotte used the same scaling arguments for $10$
earthquakes in California \cite{turcotte, shcherbakov}. Time
intervals of $92$, $183$, $365$, $730$, and $1095$ days are taken
except \c{C}anakkale-Yenice, Mu\c{s}-Varto, and Adapazar\i-Mudurnu
earthquakes. It should also be noted that for all $6$ earthquakes we
took $m_{L}$, Richter magnitudes.

For Kocaeli-G\"{o}lc\"{u}k earthquake the mainshock is $m_{ms}=7.4$
and the largest detected aftershock had a magnitude
$m_{as}^{max}=5.8$. From equation (\ref{denklem3}) the difference in
magnitude between the mainshock and largest aftershock is $\Delta
m=1.6$. We have correlated the aftershock frequency magnitude data
given in Figure 1 with G-R scaling (\ref{denklem4}) and find
$b=0.91\pm 0.05$ and $a=4.97\pm 0.20$. From equation
(\ref{denklem5}) the inferred magnitude of the largest aftershock is
$m^{*}=5.46\pm 0.37$. From equation (\ref{denklem6}) the difference
in magnitude between the mainshock and the inferred largest
aftershock is $\Delta m^{*}=1.94\pm 0.37$. We applied the same
procedure to the all $6$ earthquakes and found a, b, $m^{*}$, and
$\Delta m^{*}$ parameters.

\begin{figure}
\begin{center}
\includegraphics[width=0.75\textwidth]{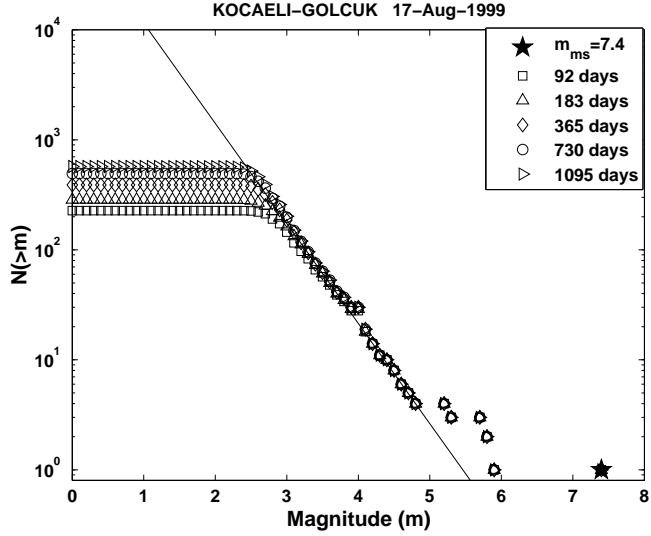} \caption{Frequency-magnitude
distribution of Kocaeli-G\"{o}lc\"{u}k earthquake}
\label{kocaeli-golcuk}
\end{center}
\end{figure}

For these $6$ earthquakes, the a, b, $m_{ms}$, $m_{as}^{max}$,
$\Delta m$, $m^{*}$, and $\Delta m^{*}$ values are given in Table
\ref{tablo4}.\\

\begin{table}
\begin{center}
\caption{Summaries of the data and results that show b and a
parameters } \vspace{0.5cm}\label{tablo4}
\begin{tabular}{|c|c|c|c|}
  \hline
  Earthquake & Date (mm/dd/yy)& b & a \\
  \hline
  \c{C}anakkale-Yenice & 03/18/53 & $0.82\pm0.07$ & $4.69\pm0.33$ \\
  \hline
  Bolu-Abant & 05/26/57 & $0.61\pm0.05$ & $3.81\pm0.24$ \\
  \hline
  Mu\c{s}-Varto & 08/19/66 & $0.95\pm0.09$ & $5.20\pm0.37$ \\
  \hline
  Adapazar\i-Mudurnu & 07/22/67 & $1.22\pm0.10$ & $6.62\pm0.45$ \\
  \hline
  Kocaeli-G\"{o}lc\"{u}k & 08/17/99 & $0.91\pm0.05$ & $4.97\pm0.20$  \\
  \hline
  D\"{u}zce & 11/12/99 & $0.80\pm0.03$ & $4.85\pm0.13$ \\
  \hline
\end{tabular}
\end{center}
\end{table}

\begin{table}
\begin{center}
\caption{Summaries of the data and results that show $m_{ms}$,
$m_{as}^{max}$, $\Delta m$, $m^{*}$, $\Delta m^{*}$ parameters}
\vspace{0.5cm}\label{tablo5}
\begin{tabular}{|c|c|c|c|c|c|}
  \hline
  Earthquake & $m_{ms}$ & $m_{as}^{max}$ & $\Delta m$ & $m^{*}$ & $\Delta m^{*}$ \\
  \hline
  \c{C}anakkale-Yenice & 7.2 & 5.4 & 1.8 & $5.72\pm0.63$ & $1.48\pm0.63$ \\
  \hline
  Bolu-Abant & 7.1 & 5.9 & 1.2 & $6.25\pm0.65$ & $0.85\pm0.65$ \\
  \hline
  Mu\c{s}-Varto & 6.9 & 5.3 & 1.6 & $5.47\pm0.65$ & $1.43\pm0.65$ \\
  \hline
  Adapazar\i-Mudurnu & 7.2 & 5.4 & 1.8 & $5.43\pm0.58$ & $1.77\pm0.58$ \\
  \hline
  Kocaeli-G\"{o}lc\"{u}k & 7.4 & 5.8 & 1.6 & $5.46\pm0.37$ & $1.94\pm0.37$ \\
  \hline
  D\"{u}zce & 7.2 & 5.4 & 1.8 & $6.06\pm0.28$ & $1.14\pm0.28$ \\
  \hline
  \end{tabular}
  \end{center}
  \end{table}

According to our results, the mean of the differences between
mainshock and largest detected aftershock magnitudes is
$\overline{\Delta m}=1.63\pm0.23$. The mean of the inferred values
of $\Delta m^{*}$ obtained from the best fit of equation
(\ref{denklem7}) is $\overline{\Delta m^{*}}=1.42\pm0.18$. In
addition for these earthquakes the mean of b values is
$\overline{b}=0.81\pm0.02$.\

\begin{figure}
\begin{center}
\includegraphics[width=0.75\textwidth]{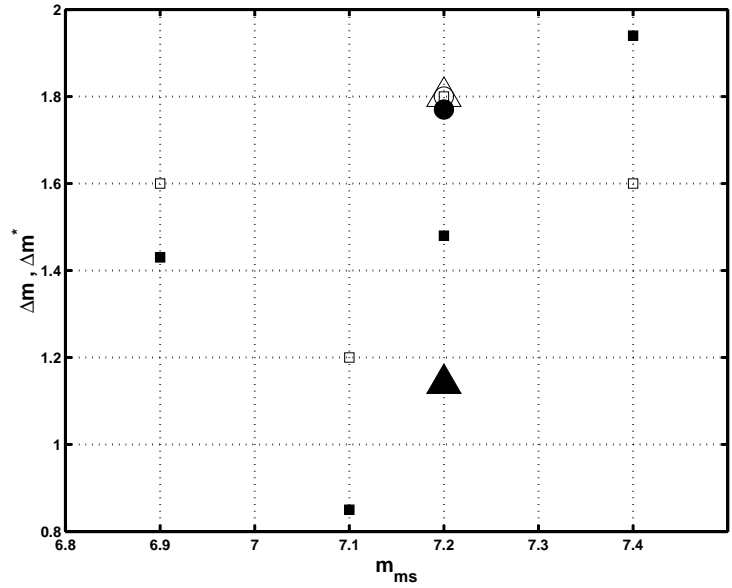} \caption{Dispersion of the
magnitude differences $\Delta m$ and $\Delta m^{*}$ on the mainshock
magnitude $m_{ms}$. White symbols correspond to
$\Delta m$ values and black symbols correspond to $\Delta m^{*}$
values.} \label{scattermakale}
\end{center}
\end{figure}

\begin{figure}
\begin{center}
\includegraphics[width=0.75\textwidth]{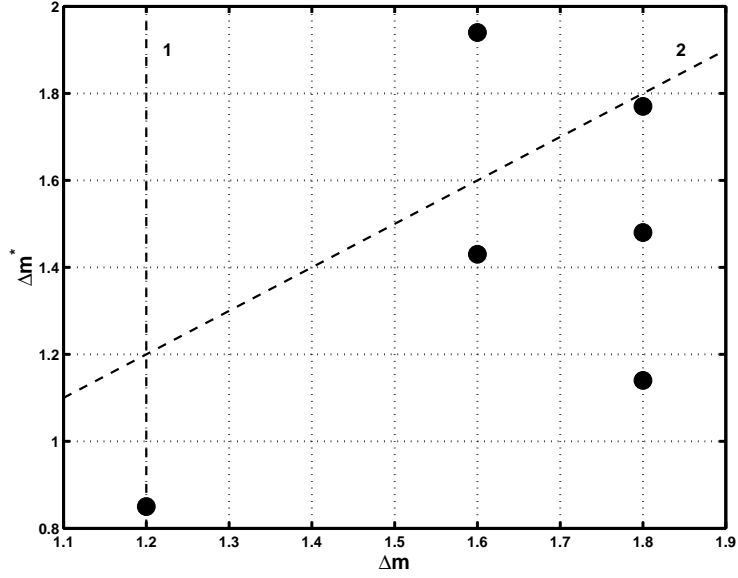} \caption{The relation between
$\Delta m$ and $\Delta m^{*}$. Line $1$ shows the harmony of our
data with the B{\aa}th's Law. Line $2$ corresponds to $y=x$ line and
shows the harmony of our data with the Modified Form of B{\aa}th's
Law.} \label{relationmakale}
\end{center}
\end{figure}

\newpage
\section{Radiated Energy During an Earthquake} Seismologists have
more recently developed a standard magnitude scale that is called
the moment magnitude, and it comes from the seismic moment. To
understand the seismic moment, we need to go back to the definition
of torque. A torque is an agent that changes the angular momentum of
a system. It is defined as the force times the distance from the
center of rotation. Earthquakes are caused by internal torques, from
the interactions of different blocks of the earth on opposite sides
of faults. It can be shown that the moment of an earthquake is
simply expressed by
\begin{equation}
M_{0}=\mu A d
\end{equation}
where $M_{0}$=Moment, $\mu$=Rock Rigidity, A=Fault Area, and
d=Slip Distance.
Both the magnitude and the seismic moment are related to the amount
of energy that is radiated by an earthquake. Radiated energy is a
particularly important aspect of earthquake behavior, because it
causes all the damage and loss of life, and additionally, it is the
greatest source of observational data. So, the seismic radiated
energy is an important physical parameter to study on earthquakes.
The relationships between the radiated energy, stress drop, and
earthquake size provides information about the physics of the
rupture process. Richter and Gutenberg, developed a relationship
between magnitude and energy. Their relationship is:
\begin{equation}\label{denklem9}
log_{10}[E(m)]=\frac{3}{2}m+11.8
\end{equation}
It should be noted that in this relation E(m) is not the total
``intrinsic" energy of the earthquake. It is only the radiated
energy from the earthquake and a small fraction of the total energy
transferred during the earthquake process. We can write this
equation in this form \cite{turcotte, utsu}.
\begin{equation}\label{denklem10}
log_{10}[E(m)]=\frac{3}{2}m+log_{10}E_{0}
\end{equation}
with $E_{0}=6.3\times 10^{4}J$. Our aim is to determine the ratio of
the total seismic energy radiated in the aftershock sequence to the
seismic energy radiated in the mainshock. This relation can be used
directly to relate the radiated energy from the mainshock $E_{ms}$
to the moment magnitude of the mainshock $m_{ms}$
\begin{equation}\label{denklem11}
E_{ms}=E_{0} 10^{(3/2) m_{ms}}
\end{equation}
\subsection{The First Calculation Method To Find The Energy Ratio
Between Mainshock and Aftershock Sequences}

The total radiated energy in the aftershock sequence $E_{as}$ is
obtained by integrating over the distributions of aftershocks
\cite{turcotte}. This can be written
\begin{equation}\label{denklem12}
E_{as}=\int_{-\infty}^{m_{as}^{max}} E(m) (-\frac{dN}{dm}) dm
\end{equation}
Taking the derivative of equation (\ref{denklem4}) with respect to
the aftershock magnitude m we have
\begin{equation}\label{denklem13}
dN= -b (ln10) 10^{a-bm} dm
\end{equation}
Putting equation (\ref{denklem13}) into equation (\ref{denklem12})
gives
\begin{equation}\label{denklem14}
E_{as}=b (ln10) 10^{a} \int_{-\infty}^{m_{as}^{max}} E(m) 10^{-bm}
dm
\end{equation}
In addition, if we turn back to equation (\ref{denklem11}) and put
it to equation (\ref{denklem14}) we get
\begin{equation}\label{denklem15}
E_{as}=b (ln10) 10^{a} E_{0}\int_{-\infty}^{m_{as}^{max}}
10^{(3/2-b)m} dm
\end{equation}
Then we take this integral and we find
\begin{equation}\label{denklem16}
E_{as}=\frac{2b}{(3-2b)}E_{0} 10^{a} 10^{(3/2-b)m_{as}^{max}}
\end{equation}
To find the ratio of the total radiated energy in aftershocks
$E_{as}$ to the radiated energy in the mainshock $E_{ms}$, we divide
equation (\ref{denklem16}) to equation (\ref{denklem11}). Then we
get the result
\begin{equation}\label{denklem17}
\frac{E_{as}}{E_{ms}}=\frac{2b}{(3-2b)} 10^{a} 10^{-bm_{as}^{max}}
10^{-3/2(m_{ms}-m_{as}^{max})}
\end{equation}
We know that $\Delta m= m_{ms}-m_{as}^{max}$ so equation
(\ref{denklem17}) takes this form:
\begin{equation}\label{denklem18}
\frac{E_{as}}{E_{ms}}=\frac{2b}{(3-2b)} 10^{a} 10^{-bm_{as}^{max}}
10^{-3/2(\Delta m)}
\end{equation}
From the equation (\ref{denklem18}), the fraction of the total
energy associated with aftershocks is given by
\begin{equation}\label{denklem19}
\frac{E_{as}}{E_{ms}+E_{as}}=\frac{1}{1+ (\frac{3-2b}{2b})
10^{3/2\Delta m} 10^{-(a-bm_{as}^{max})}}
\end{equation}
For the $6$ earthquakes considered in the previous section we had
put the b, a, $\Delta m$ and $m_{as}^{max}$ values to equation
(\ref{denklem19}) individually. Our aim is to find
$\frac{E_{as}}{E_{ms}+E_{as}}$  values for $6$
earthquakes. \\

\begin{table}
\begin{center}
\caption{Summaries of the data and results for energy values that
were taken from the first energy calculation method}
\vspace{0.5cm}\label{tablo7}
\begin{tabular}{|c|c|c|c|c|c|}
  \hline
  Earthquake & a & b & $m_{as}^{max}$ & $\Delta m$ & $\frac{E_{as}}{E_{ms}+E_{as}}$ \\
  \hline
  \c{C}anakkale-Yenice & $4.69\pm0.33$ & $0.82\pm0.07$ & 5.4 & 1.8 & 0.004 \\
  \hline
  Bolu-Abant & $3.81\pm0.24$ & $0.61\pm0.05$ & 5.9 & 1.2 & 0.017 \\
  \hline
  Mu\c{s}-Varto & $5.20\pm0.37$ & $0.95\pm0.09$ & 5.3 & 1.6 & 0.010 \\
  \hline
  Adapazar\i-Mudurnu & $6.62\pm0.45$ & $1.22\pm0.10$ & 5.4 & 1.8 & 0.009 \\
  \hline
  Kocaeli-G\"{o}lc\"{u}k & $4.97\pm0.20$ & $0.91\pm0.05$ & 5.8 & 1.6 & 0.002 \\
  \hline
  D\"{u}zce & $4.85\pm0.13$ & $0.80\pm0.03$ & 5.4 & 1.8 & 0.008 \\
  \hline
\end{tabular}
\end{center}
\end{table}

According to our results, we find the mean energy
$\frac{E_{as}}{E_{ms}+E_{as}}=0.008$ with a standard deviation
$\sigma_{\overline{E}}=0.005$. Consequently, we find that for these
earthquakes on average about $99.2$ per cent of the available
elastic energy is released during the mainshock and about $0.8$ per
cent of energy is released during the aftershocks.

\subsection{The Second Calculation Method To Find The Energy Ratio
Between Mainshock and Aftershock Sequences}

Additionally, from the study of Shcherbakov and Turcotte in $2004$
\cite{turcotte, shcherbakov}, we may derive the same energy ratio in
terms of b and $\Delta m^{*}$ values.\

The total radiated energy in the aftershock sequence $E_{as}$ is
obtained by integrating over the distributions of aftershocks
\cite{turcotte}. This can be written
\begin{equation}\label{denklem20}
E_{as}=\int_{-\infty}^{m^{*}} E(m) (-\frac{dN}{dm}) dm
\end{equation}
Taking the derivative of equation (\ref{denklem7})with respect to
the aftershock magnitude m we have
\begin{equation}\label{denklem21}
dN= -b (ln10) 10^{b(m_{ms}-\Delta m^{*}-m)} dm
\end{equation}
Putting equation (\ref{denklem21}) into equation (\ref{denklem20})
gives
\begin{equation}\label{denklem22}
E_{as}=b (ln10) 10^{b(m_{ms}-\Delta m^{*})} \int_{-\infty}^{m^{*}}
E(m) 10^{-bm} dm
\end{equation}
In addition, if we turn back to equation (\ref{denklem11}) and put
it to equation (\ref{denklem22}) we get
\begin{equation}\label{denklem23}
E_{as}=b (ln10) 10^{b(m_{ms}-\Delta m^{*})}
E_{0}\int_{-\infty}^{m^{*}}  10^{(3/2-b)m}  dm
\end{equation}
Then we take this integral and we find
\begin{equation}\label{denklem24}
E_{as}=\frac{2b}{(3-2b)}E_{0} 10^{(3/2-b)m^{*}} 10^{b(m_{ms}-\Delta
m^{*})}
\end{equation}
Using equation (\ref{denklem6}) we find
\begin{equation}\label{denklem25}
E_{as}=\frac{2b}{(3-2b)}E_{0} 10^{3/2(m_{ms}-\Delta m^{*})}
\end{equation}
To find the ratio of the total radiated energy in aftershocks
$E_{as}$ to the radiated energy in the mainshock $E_{ms}$,  we
divide equation (\ref{denklem25}) to equation (\ref{denklem11}).
Then we get the result
\begin{equation}\label{denklem26}
\frac{E_{as}}{E_{ms}}=\frac{2b}{(3-2b)} 10^{-3/2\Delta m^{*}}
\end{equation}
If we further assume that all earthquakes have the same seismic
efficiency (ratio of radiated energy to the total drop in stored
elastic energy), then this ratio is also the ratio of the drop in
stored elastic energy due to the aftershocks to the drop in stored
elastic energy due to the mainshock. From equation (\ref{denklem26})
the fraction of the total energy associated with aftershocks is
given by

\begin{equation}\label{denklem27}
\frac{E_{as}}{E_{ms}+E_{as}}=\frac{1}{1+\frac{3-2b}{2b} 10^{3/2
\Delta m^{*}}}
\end{equation}

For the $6$ earthquakes considered in the previous section we had
put the b and  $m^{*}$ values to equation (\ref{denklem27})
individually. Our aim is to find $\frac{E_{as}}{E_{ms}+E_{as}}$
values for $6$ earthquakes considered. The obtained results are
summarized in Table (\ref{tablo8}).
\begin{table}
\begin{center}
\caption{Summaries of the data and results for energy values that
were taken from the second energy calculation method}
\vspace{0.5cm}\label{tablo8}
\begin{tabular}{|c|c|c|c|}
  \hline
  Earthquake & b & $\Delta m^{*}$ & $\frac{E_{as}}{E_{ms}+E_{as}}$ \\
  \hline
  \c{C}anakkale-Yenice  & $0.82\pm0.07$ & $1.48\pm0.63$ & 0.007 \\
  \hline
  Bolu-Abant & $0.61\pm0.05$ & $0.85\pm0.65$ & 0.035 \\
  \hline
  Mu\c{s}-Varto & $0.95\pm0.09$ & $1.43\pm0.65$ & 0.012 \\
  \hline
  Adapazar\i-Mudurnu & $1.22\pm0.10$ & $1.77\pm0.57$ & 0.010 \\
  \hline
  Kocaeli-G\"{o}lc\"{u}k & $0.91\pm0.05$ & $1.94\pm0.37$ & 0.002 \\
  \hline
  D\"{u}zce & $0.80\pm0.03$ & $1.14\pm0.28$ & 0.022 \\
  \hline
\end{tabular}
\end{center}
\end{table}

According to our results, we find the mean energy
$\frac{E_{as}}{E_{ms}+E_{as}}=0.015$ with a standard deviation
$\sigma_{\overline{E}}=0.012$. Consequently, we find that the ratio
of radiated energy in aftershocks to the radiated energy in the
mainshock is constant. This is consistent with the generally
accepted condition of self-similarity for earthquakes. For these
earthquakes on average about $98.5$ per cent of the available
elastic energy goes into the mainshock and about $1.5$ per cent into
the aftershocks.

\section{Discussion}

Earthquakes occur in clusters. After one earthquake happens, we
usually see others at nearby or identical location. Clustering of
earthquakes usually occurs near the location of the mainshock. The
stress on the mainshock's fault changes drastically during the
mainshock and that fault produces most of the aftershocks. This
causes a change in the regional stress, the size of which decreases
rapidly with distance from the mainshock. Sometimes the change in
stress caused by the mainshock is great enough to trigger
aftershocks on other, nearby faults. It is accepted that aftershocks
are caused by stress transfer during an earthquake. When an
earthquake occurs there are adjacent regions where the stress is
increased. The relaxation of these stresses causes aftershocks
\cite{turcotte, rybicki, das, mendoza, king, marcellini,
hardebeck}.\

Several scaling laws are also found to be universally valid for
aftershocks \cite{kisslinger, turcotte, shcherbakov}. These are:
\begin{enumerate}

\item  Gutenberg-Richter frequency-magnitude scaling
\item  The modified Omori's law for the temporal decay of aftershocks
\item  B{\aa}th's law for the magnitude of the largest
aftershock\
\end{enumerate}

In this work we used both B{\aa}th's law and G-R scaling. Our
aim is to find an upper cutoff magnitude $m^{*}$ for a given
aftershock sequence. Using relation (\ref{denklem5}), we get related
a and b values in the G-R scaling. B{\aa}th's law states that, to a
good approximation, the difference in magnitude between mainshock
and its largest aftershock is a constant independent of the
mainshock magnitude.\ A modified form of B{\aa}th's law was proposed
by Shcherbakov and Turcotte in $2004$ \cite{turcotte, shcherbakov}.
They considered $10$ large earthquakes that occurred in California
between $1987$ and $2003$ with magnitudes equal to or greater than
$m_{ms}\geq5.5$. According to their theory the mean difference in
magnitudes between these mainshocks and their largest detected
aftershocks is $1.16\pm0.46$. This result is consistent with
B{\aa}th's Law. They found the mean difference in magnitudes between
the mainshocks and their largest inferred aftershocks is
$1.11\pm0.29$. They also calculated the partitioning of energy
during a mainshock-aftershock sequence and found that about $96$ per
cent of the energy dissipated in a sequence is associated with the
mainshock and the rest ($4$ per cent) is due to aftershocks. Their
results are given in Table \ref{tablo9}. We applied the Modified
Form of B{\aa}th's Law to our $6$ large earthquakes that occurred on
the North Anatolian Fault Zone (NAFZ) in Turkey. We followed the
same calculation process.

\begin{table}
\begin{center}
\caption{Comparision the results} \vspace{0.5cm}\label{tablo9}
\begin{tabular}{|c|c|c|}
  \hline
Parameters & Turcotte and Shcherbakov & Kurnaz and Yalcin \\
  \hline
  $\overline{\Delta m}$ & $1.16\pm0.46$ & $1.63\pm0.23$  \\
  \hline
  $\overline{\Delta m^{*}}$ & $1.11\pm0.29$ & $1.42\pm0.18$ \\
  \hline
  $\frac{E_{as}}{E_{ms}+E_{as}}$ &  $0.038$ &  $0.015$ \\
  \hline
\end{tabular}
\end{center}
\end{table}

According to Table \ref{tablo9}, for the North Anatolian Fault Zone
(NAFZ), a large fraction of the accumulated energy is released in
the mainshock and only a relatively small fraction of the
accumulated energy is released in the aftershock sequence. The
results of Turcotte and Shcherbakov are for the ten earthquakes in
California on the San Andreas Fault Zone. Although SAFZ (in
California) and NAFZ (in Turkey) have the same seismic properties,
the released energy during the mainshocks in the NAFZ is much
greater than the released energy during the mainshocks in the SAFZ.

Figure 7 shows the dispersion of the magnitude differences $\Delta
m$ and $\Delta m^{*}$ on the mainshock magnitude $m_{ms}$. In this
figure, white symbols correspond to $\Delta m$ values and black
symbols correspond to $\Delta m^{*}$ values. Square, circle and
triangle were used to prevent the coincides of the data on the
figure; because for six earthquakes, three of them have the same
mainshock magnitude $m_{ms}=7.2$.

Additionally, Figure 8 gives us the relation between $\Delta m$ and
$\Delta m^{*}$. In this figure, line $1$ shows the harmony of our
data with the B{\aa}th's Law. According to Figure 8, our data do not
show harmony with the B{\aa}th's Law. B{\aa}th's Law states that the
difference in magnitude between a mainshock and its largest detected
aftershock is constant, regardless of the mainshock magnitude and it
is about $1.2$ \cite{turcotte, shcherbakov, sornette, bath}. But in
Figure 8, only one earthquake has $\Delta m$ values equal to $1.2$.
The other five earthquakes have $\Delta m$ values greater than
$1.2$. Consequently, only $17$ per cent of our data show harmony
with the B{\aa}th's Law. The rest part ( $83$ per cent) of our data
do not show harmony with the B{\aa}th's Law.\

The constancy of the differences in magnitudes between mainshocks
and their largest aftershocks is an indication of scale-invariant
behavior of aftershock sequences.\

In Figure 8, line $2$ shows the harmony of our data with the
Modified Form of B{\aa}th's Law. This line corresponds to $y=x$
line. If the Modified Form of B{\aa}th's Law gave us perfect
results, $\Delta m$ and $\Delta m^{*}$ values would be close to each
other along this line. Hence, they would be the near of line $2$.
But in Figure 8, only one earthquake takes place on the upper side
of this line. The remaining five earthquakes take place on the lower
side of this line. Consequently, our data do not show harmony with
the Modified Form of B{\aa}th's Law.\

The other important conclusion is that we know  most of the energy
is released during the mainshock. Therefore, after the mainshock the
community and government may begin their work to rescue people from
the debris without wasting any time.

\ack We thank Professor Niyazi T\"{u}rkelli for his help and many
helpful advices. The KOERI Data were provided by Bo\u{g}azi\c{c}i
University Kandilli Observatory and Earthquake Research Institute.


\section*{References}

\begin{thebibliography}{10}
\expandafter\ifx\csname bibnamefont\endcsname\relax
  \def\bibnamefont#1{#1}\fi
\expandafter\ifx\csname bibfnamefont\endcsname\relax
  \def\bibfnamefont#1{#1}\fi
\expandafter\ifx\csname url\endcsname\relax
  \def\url#1{\texttt{#1}}\fi
\expandafter\ifx\csname urlprefix\endcsname\relax\def\urlprefix{URL
}\fi \providecommand{\bibinfo}[2]{#2}
\providecommand{\eprint}[2][]{\url{#2}}

\bibitem{kanamori}
\bibinfo{author}{\bibfnamefont{H.} \bibnamefont{Kanamori}},
  \bibinfo{author}{\bibfnamefont{Emily E.} \bibnamefont{Brodsky}},
  \bibinfo{journal}{Reports on Progress in Physics} \textbf{\bibinfo{volume}{67}},
pp\bibinfo{pages}{1429-1496} (\bibinfo{year}{2004}).

\bibitem{kisslinger}
\bibinfo{author}{\bibfnamefont{C.} \bibnamefont{ Kisslinger}},
  \bibinfo{journal}{Advances in Geophysics} \textbf{\bibinfo{volume}{38}},
pp\bibinfo{pages}{1-36} (\bibinfo{year}{1996}).

\bibitem{turcotte}
\bibinfo{author}{\bibfnamefont{R.} \bibnamefont{Shcherbakov}},
  \bibinfo{author}{\bibfnamefont{Donald L.} \bibnamefont{Turcotte}},
   \bibinfo{journal}{Bulletin of the Seismological Society of America}
\textbf{\bibinfo{volume}{94}},
 pp\bibinfo{pages}{1968-1975} (\bibinfo{year}{2004}).

\bibitem{shcherbakov}
\bibinfo{author}{\bibfnamefont{R.} \bibnamefont{Shcherbakov}},
  \bibinfo{author}{\bibfnamefont{Donald L.} \bibnamefont{Turcotte}},
   \bibinfo{author}{\bibfnamefont{John B.} \bibnamefont{Rundle}},
   \bibinfo{journal}{Pure and Applied Geophysics}
\textbf{\bibinfo{volume}{162}},
 pp\bibinfo{pages}{1051-1076} (\bibinfo{year}{2005}).

\bibitem{gutenberg}
\bibinfo{author}{\bibfnamefont{B.}~\bibnamefont{ Gutenberg}},
  \bibinfo{author}{\bibfnamefont{C. F.} \bibnamefont{Richter}}
   \emph{\bibinfo{title}{Seismicity of the Earth and Associated Phenomena}}
    (\bibinfo{publisher}{Princeton Univ. Press,  Princeton, New Jersey},
  \bibinfo{year}{1954}).

\bibitem{omori}
\bibinfo{author}{\bibfnamefont{F.} \bibnamefont{Omori}},
  \bibinfo{journal}{Journal of College of Science of the Imperial University of Tokyo}
\textbf{\bibinfo{volume}{7}},
  \bibinfo{pages}{111-200} (\bibinfo{year}{1894}).

\bibitem{sornette}
\bibinfo{author}{\bibfnamefont{A.} \bibnamefont{Helmstetter}},
  \bibinfo{author}{\bibfnamefont{D.} \bibnamefont{Sornette}},
  \bibinfo{journal}{Geophysical Research Letters}
\textbf{\bibinfo{volume}{30}},
  \bibinfo{doi}{10.1029/2003GL018186} (\bibinfo{year}{2003}).

\bibitem{knopoff}
\bibinfo{author}{\bibfnamefont{Y. Y.} \bibnamefont{Kagan}},
  \bibinfo{author}{\bibfnamefont{L.} \bibnamefont{Knopoff}},
  \bibinfo{journal}{Journal of  Geophysical Research}
\textbf{\bibinfo{volume}{86}},
  \bibinfo{pages}{2853-2862} (\bibinfo{year}{1981}).

\bibitem{ogata}
\bibinfo{author}{\bibfnamefont{Y.} \bibnamefont{Ogata}},
  \bibinfo{journal}{Journal of the American Statistical Association}
\textbf{\bibinfo{volume}{83}},
  \bibinfo{pages}{9-27} (\bibinfo{year}{1988}).

\bibitem{bath}
\bibinfo{author}{\bibfnamefont{M.} \bibnamefont{Bath}},
  \bibinfo{journal}{Tectonophysics} \textbf{\bibinfo{volume}{2}},
  \bibinfo{pages}{483-514} (\bibinfo{year}{1965}).

\bibitem{frolich}
\bibinfo{author}{\bibfnamefont{C.} \bibnamefont{Frolich}},
  \bibinfo{author}{\bibfnamefont{S. D.} \bibnamefont{Davis}},
  \bibinfo{journal}{Journal of Geophysical Research} \textbf{\bibinfo{volume}{98}},
  \bibinfo{pages}{631-644} (\bibinfo{year}{1993}).

\bibitem{kandilli}
\bibinfo{author}{\bibnamefont{Bogazici University Kandilli Observatory and Earthquake Research Institute}},
  \bibinfo{journal}{http://www.koeri.boun.edu.tr, 2006}.

\bibitem{molchan}
\bibinfo{author}{\bibfnamefont{G. M.} \bibnamefont{Molchan}},
  \bibinfo{author}{\bibfnamefont{O. E.} \bibnamefont{Dmitrieva}},
  \bibinfo{journal}{Geophysical Journal International}
\textbf{\bibinfo{volume}{109}},
  \bibinfo{pages}{501-516} (\bibinfo{year}{1992}).

\bibitem{kagan}
\bibinfo{author}{\bibfnamefont{Y. Y.} \bibnamefont{Kagan}},
  \bibinfo{journal}{Bulletin of Seismological Society of America}
\textbf{\bibinfo{volume}{92}},
  \bibinfo{pages}{641-655} (\bibinfo{year}{2002}).

\bibitem{utsu}
\bibinfo{author}{\bibfnamefont{T.}~\bibnamefont{Utsu}},
   \emph{\bibinfo{title}{Relationship between magnitude scales}}
    (\bibinfo{publisher}{International Handbook of Earthquake and Engineering Seismology, W. H. K.},
     \bibinfo{year}{2002}).

\bibitem{rybicki}
\bibinfo{author}{\bibfnamefont{K.} \bibnamefont{Rybicki}},
  \bibinfo{journal}{Physics of the Earth and Planetary Interiors}
\textbf{\bibinfo{volume}{7}},
  \bibinfo{pages}{409-422} (\bibinfo{year}{1973}).

\bibitem{das}
\bibinfo{author}{\bibfnamefont{S.} \bibnamefont{Das}},
  \bibinfo{author}{\bibfnamefont{C. H.} \bibnamefont{Scholz}},
  \bibinfo{journal}{Bulletin of Seismological Society of America}
\textbf{\bibinfo{volume}{71}},
  \bibinfo{pages}{1669-1675} (\bibinfo{year}{1981}).

\bibitem{mendoza}
\bibinfo{author}{\bibfnamefont{C.} \bibnamefont{Mendoza}},
  \bibinfo{author}{\bibfnamefont{S. H.} \bibnamefont{Hartzell}},
  \bibinfo{journal}{Bulletin of Seismological Society of America}
\textbf{\bibinfo{volume}{78}},
  \bibinfo{pages}{1438-1449} (\bibinfo{year}{1988}).

\bibitem{king}
\bibinfo{author}{\bibfnamefont{G. C. P.} \bibnamefont{King}},
   \bibinfo{author}{\bibfnamefont{R. S.} \bibnamefont{Stein}},
    \bibinfo{author}{\bibfnamefont{J.} \bibnamefont{Lin}},
  \bibinfo{journal}{Bulletin of Seismological Society of America}
\textbf{\bibinfo{volume}{84}},
  \bibinfo{pages}{935-953} (\bibinfo{year}{1994}).

\bibitem{marcellini}
\bibinfo{author}{\bibfnamefont{A.} \bibnamefont{Marcellini}},
  \bibinfo{journal}{Journal of  Geophysical Research}
\textbf{\bibinfo{volume}{100}},
  \bibinfo{pages}{6463-6468} (\bibinfo{year}{1995}).

\bibitem{hardebeck}
\bibinfo{author}{\bibfnamefont{J. L.} \bibnamefont{Hardebeck}},
  \bibinfo{author}{\bibfnamefont{J. J.} \bibnamefont{Nazareth}},
   \bibinfo{author}{\bibfnamefont{E.} \bibnamefont{Hauksson}},
  \bibinfo{journal}{Journal of  Geophysical Research} \textbf{\bibinfo{volume}{103}},
  \bibinfo{pages}{24,427-24,437} (\bibinfo{year}{1998}).

\end{thebibliography}
\end{document}